\documentclass{article}

\usepackage{arxiv}

\usepackage[utf8]{inputenc} 
\usepackage[T1]{fontenc}    
\usepackage{hyperref}       
\usepackage{url}            
\usepackage{booktabs}       
\usepackage{amsfonts}       
\usepackage{nicefrac}       
\usepackage{microtype}      
\usepackage{lipsum}		
\usepackage{graphicx}

\title{Gather - a better way to codehack online}

\date{}

\author{ \href{https://orcid.org/0000-0002-0672-833X}{\includegraphics[scale=0.06]{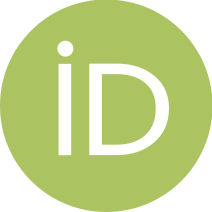}\hspace{1mm}Rika Kobayashi\thanks{Corresponding author}}\\
1. Leonard Huxley Building 56\\
Australian National University\\
Canberra, Australia\\
2. University of Queensland\\
Brisbane, Australia\\
	\texttt{Rika.Kobayashi@anu.edu.au} \\
	\And
	\href{https://orcid.org/0000-0002-6711-6345}{\includegraphics[scale=0.06]{orcid.png}\hspace{1mm}Sarah Jaffa$^*$} \\
Advanced Research Computing Centre\\
University College London\\
London, United Kingdom\\
	\texttt{s.jaffa@ucl.ac.uk} \\
	\And
	Jiachen Dong \\
	University of Queensland\\
	Brisbane, Australia \\
	\And
	\href{https://orcid.org/0000-0003-2868-7994}{\includegraphics[scale=0.06]{orcid.png}\hspace{1mm}Roger D. Amos} \\
	University of Technology Sydney\\
	Sydney, Australia\\
	\And
	\href{https://orcid.org/0000-0003-4312-2537}{\includegraphics[scale=0.06]{orcid.png}\hspace{1mm}Jeremy Cohen} \\
    Department of Computing\\
	Imperial College London\\
	London, United Kingdom\\
	\And
	\href{https://orcid.org/0000-0002-0011-6922}{\includegraphics[scale=0.06]{orcid.png}\hspace{1mm}Emily F. Kerrison} \\
	Sydney Institute for Astronomy\\
	University of Sydney\\
	Sydney, Australia \\
}

\renewcommand{\shorttitle}{Gather - a better way to codehack online}

\begin{document}
\maketitle

\begin{abstract}
A virtual hands-on computer laboratory has been designed within the Gather online meeting platform. Gather's features such as spatial audio, private spaces and interactable objects offer scope for great improvements over currently used platforms, especially for small-group based teaching. We describe our experience using this virtual computer laboratory for a recent 'Python for Beginners' workshop held as part of the Software Sustainability Institute's 2022 Research Software Camp.
\end{abstract}


\section{INTRODUCTION}
The hands-on computer laboratory remains one of the relatively little-addressed challenges of remote teaching \cite{JChemEd}. The main purpose of the in-person equivalent is interactivity, often described as active learning, where typically an instructor delivers a course and the students then work through exercises on individual computers with a group of tutors overseeing their work \cite{Nikolic}. This small group activity is also more generally used in brainstorming  workshops or hackathons. Since the outbreak of the COVID-19 pandemic most such teaching has been carried out predominantly using common methods for online teaching such as through video recordings or conference calls. This delivery method can be described as one-to-many communication; effectively only one conversation can be carried out at a time. The more individual interactions that can underpin small group work often leverage breakout rooms, a web-conferencing feature that allows sub-groups to “break out” into smaller meetings from within the main meeting \cite{Stanford}. However these can feel isolating, with connection to the main room being lost, and in environments where the workshop host controls movement between the rooms, who and when, a sense of being "sent to Coventry".
Gather \cite{Gather}, a virtual meeting space founded in 2020 with the aim of providing a mechanism for friends to stay connected, has many features that make it attractive for hosting hands-on workshops: its use of spatial audio and private spaces, and the ability to embed interactive objects.There have now been many examples of Gather being used as a platform within the academic community to host conference sessions, poster sessions or as a platform to support the social elements of hybrid or online events where participants may be missing out on the benefits of being able to interact in person with colleagues, collaborators or other event attendees. In particular, it avoids many of the communication issues involved with breakout rooms \cite{Stanford}. Spatial audio is an audio setup to simulate surround-sound, generally associated with 3D Virtual Reality to give a sense of immersion. In Gather, a 2D virtual platform, it is simplified to a volume control determined by the proximity of participants represented by an avatar. Thus it is possible to have many-to-many conversations in the same space without disturbing others. Furthermore, Gather’s spotlight feature allows one-to-many broadcast to the whole room, ideal for teaching. Private spaces are areas within the room, where only users in that space can see and hear each other, regardless of their proximity. Likewise, nearby participants who are outside a private space cannot hear conversations within the private space unless they enter the space themselves. Private spaces can be thought of as {\it de facto}  breakout rooms, but where the students remain in the main room and can carry out side-conversations, and so remain connected. Furthermore, the students retain autonomy - they get to choose when they join the breakout discussions and when they leave.  A further refinement to the spatial audio that proved invaluable was Gather’s bubbles, described as being for “whispering”, which allowed tutors to give one-on-one tuition when the need arose.
Gather’s greatest strength, however, is its ability to embed a wealth of interactive objects, such as video, shared documents, games, and most importantly for this workshop, a python coding platform.
In this paper we describe our experience designing and delivering a “Python for Beginners“ hands-on workshop within the Gather platform as part of the Software Sustainability Institute's  Research Software Camp 2022 \cite{RSCamp}.

\section{DELIVERY OF A BEGINNER-LEVEL PYTHON COURSE}
The Software Sustainability Institute~\cite{SSI} is based within the UK at the Universities of Edinburgh, Manchester, Oxford and Southampton. The institute was started in 2010 and its activities are most simply summarised by its motto: ``\textit{Better Software, Better Research}``~\cite{Goble14}. Its motivation is to promote best practice for scientific computing \cite{Wilson14} and to this end it has been running twice-yearly ``Research Software Camps``. Each Research Software Camp runs for approximately two weeks and encompasses a range of events and activities under a particular theme. As part of the ``\textit{Next steps in coding}`` research software camp held in May 2022, it was decided to run a beginner-level online Python course. The team leading this course were keen to trial the use of Gather as an alternative, and potentially more suitable, platform for online training when compared to the video conferencing platforms generally used for online training in the research software community over recent years.

The course was based on material from The Carpentries~\cite{TheCarpentries}, an organisation providing a wide range of open source training materials covering core research computing skills to support the software, data and libraries communities. Carpentries materials are community developed and maintained and core lessons, such as the two introductory Python lessons that are available under the Software Carpentry programme, have been taught and optimised over several years to offer a high-quality set of training material. The Carpentries operate programmes for training instructors in the core pedagogical skills and teaching methods to help ensure that materials taught as part of officially badged Carpentries workshops are delivered according to teaching best practices and that learners receive the best possible training experience.

While the course being described here represents only one element of a full Software Carpentry series and was not, therefore, being operated as an official Carpentries workshop, several of the members of the course organising team and helpers are experienced with running Carpentries courses and some are trained Carpentries instructors. We also had a representative from the Carpentries involved with the course.

Why teach Python as part of this trial of using Gather for online training? Python is now, perhaps, the most widely used programming language in the research community. The high-level nature of the language means it offers a lower barrier for beginners to learn when compared with low-level (or lower-level) languages such as C/C++, Java, or FORTRAN. It also provides a huge ecosystem of software libraries and packages that can support the development of software for almost any imaginable use case. This flexibility, relative ease of learning and its existing popularity within the research community make Python a great choice for researchers taking their first steps in coding, or looking to learn another language.

When learning a new programming language, there is no substitute for hands-on experience. This is especially true when undertaking the first steps with a new language. Simply listening to an instructor reading through (or even demonstrating) language syntax, structures and concepts has much less value than being able to follow along yourself, make mistakes, investigate and correct them and grow your understanding as part of a hands-on, practical process. Carpentries material is designed to specifically support this mode of teaching with lots of examples and exercises embedded into the material and a structure that supports learners following along with the instructor as they work through the material. This `live coding' approach lends itself well to a classroom environment where there are helpers available to answer questions and it's easy to look at a participant's screen and spot quickly whether they've simply made a typing error in a command, or whether they are looking for further explanation of a concept or exercise. In the following section we'll look in more detail about the specific challenges of technical skills training in an online environment. There is further discussion about this specific workshop and our experiences of undertaking training within the Gather platform in the discussion in Section~\ref{section:discussion}.

\section{DESIGNING THE TECHNICAL STACK}
\begin{figure}[h]
  \includegraphics[width=\linewidth]{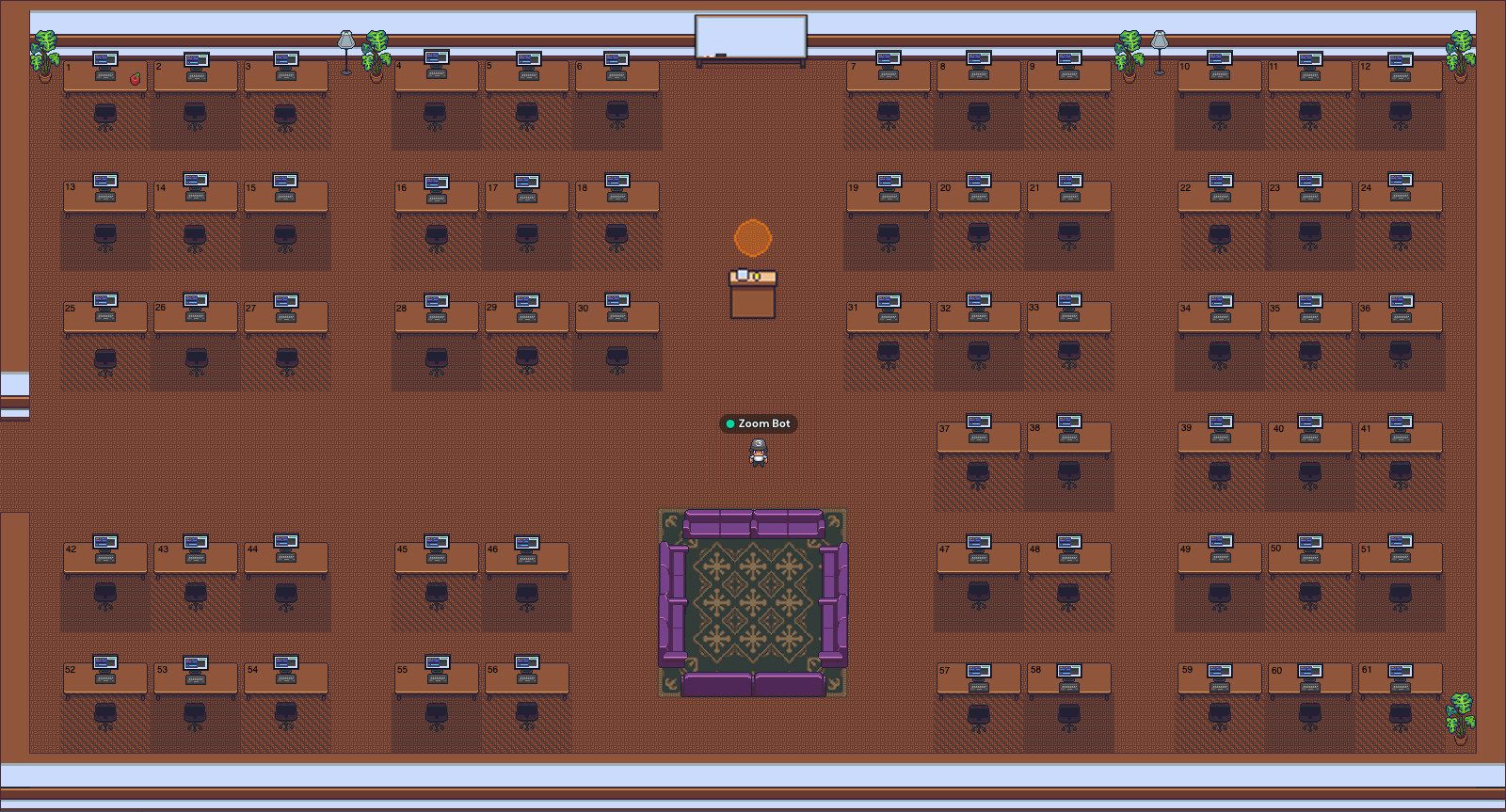}
  \caption{Gather Computer Laboratory}
  \label{fig:GatherLab}
\end{figure}

The starting point of designing the Gather space for the workshop was to reproduce the in-person layout of a typical teaching computer laboratory: rows of desks with computers, a lectern for the instructor and a whiteboard (see Fig.~\ref{fig:GatherLab}).

There are many studies in the literature regarding optimum class sizes for online teaching, see e.g. Ref. \cite{Taft} and references therein, and there is no definitive answer depending on the type of class – its composition, level, and purpose. There are fewer such studies concerning computer laboratories \cite{Nikolic,Akinola} so the number of participants was limited to 60, mainly determined by comfort level from previous experiences in delivering workshops both in-person and online. Similarly, a tutor student ratio of 1 to 10 was chosen on this basis. 
Most of the studies, however, agreed on the importance of building a sense of community and keeping a sense of connectedness, so to encourage peer interaction we grouped banks of three computer desks into a Gather private space, allowing students to talk to their neighbours in their row without disturbing the rest of the room.
Despite programming courses being grounded in computers, online computer teaching laboratories are not as straightforward to set up as would be imagined due to a variety of factors: software issues, technology, and classroom management \cite{JChemEd}.
For the virtual computers, choice of coding platform was determined by considering the following desirable features:
\begin{itemize}
\item[--]	Ease of use: the platform should be self-contained requiring minimal setup. Students should not need to be system administrators, and installation of the python programming language and associated packages, especially through conda installs, can take up a lot of disk space.
\item[--] Account creation: user management is time-consuming. Participants are reluctant to sign up to too many platforms.
\item[--] Embeddable vs shared links: a requirement specific to Gather, but related to participants wishing to stay within the one platform and not having to negotiate multiple browser windows.
\item[--] Connectivity/Usability: as with any online platform bandwidth limitations will affect interactivity which is an important part of hands-on teaching.
\item[--] Jupyter notebook format: many shared python scripts are in the form of Jupyter notebooks so it was felt students would be more comfortable with this format.
\end{itemize}

\pagebreak 
The available free online coding platforms supporting python coding, fulfilling most of these criteria, that were considered were:
\begin{itemize}
\item[--] In-house Jupyter Lab \cite{OOD}
\item[--] CoCalc \cite{CoCalc}
\item[--] REPL \cite{REPL}
\item[--] Binder \cite{Binder}
\item[--] Jupyter Lite \cite{JupLite}
\item[--] Google Colab \cite{Colab}
\end{itemize}

\begin{table}[h]
  \caption{Desired features for the online python platform}
  \label{tab:eval}
\begin{minipage}{\textwidth}
  \begin{tabular}{lccccc}
    \toprule
    &Account creation&Embeddable&Self-contained&Usability &Jupyter\\
    \midrule
    In-house Jupyter Lab & Yes & No & Yes\footnote{Prior work needed to provide the required python packages} & Poor&Yes\\
    Cocalc & Yes & No& No & Very poor & Yes\\
    REPL & ???& Yes & ??? & Good & No \\
    Binder & ???  & No\footnote{Should work in theory}& ??? & Good & Yes \\
    Jupyter Lite & No & Yes & ???& Very Good& Yes \\
    Google Colab & Yes\footnote{Requires Google account} & No & Yes& Very good & Yes \\
  \bottomrule
\end{tabular}
\end{minipage}
\end{table}

Note: that there exist many commercial offerings that are now widely used for online training, most notably AWS Educate, Google Cloud, Microsoft Azure, but these all involve payment.

Of the platforms evaluated in Table~\ref{tab:eval}, Cocalc and the in-house Jupyter Lab quickly ruled themselves out, not just on the need for account creation but on the basis of performance. The free tier version of Cocalc displayed severe connectivity issues so that it was impossible to do much. The in-house Jupyter Lab, based at Australia's National Computational Infrastructure \cite{OOD}, had no connection problems but was unable to run a sample Jupyter notebook to completion consistently. This could be due to the shared nature of the infrastructure. This infrastructure has since been upgraded and so may no longer be a problem.
The first functionality to fulfil was chosen to be embedability. For this REPL and Jupyter Lite embed seamlessly into the Gather computers. In theory, Binder should be able to embed but in reality it could only do so partially, losing connection on generating the workbook. Google Colab, despite leveraging Google Drive, only allows shared links. This is a design decision of Google as Google Docs and Google Sheets can be embedded. However, despite the ease of use of embedding in REPL and Jupyter Lite, it was discovered that when navigating away from the workbook e.g. to watch the lecture, re-entering the workbook reset it, requiring the rediscovering and reloading of the notebook each time. This left Google Colab as the chosen candidate. A link to Google Colab was embedded into the virtual computer which opened a Jupyter Notebook in a new web browser tab. This turned out not to be as inconvenient as first thought. Not only could tutors see the Notebook simultaneously without the need to screenshare but there was an unexpected side-benefit of using Google Drive to share the datasets needed for the workshop. 

\begin{figure}[h]
  \includegraphics[width=\linewidth]{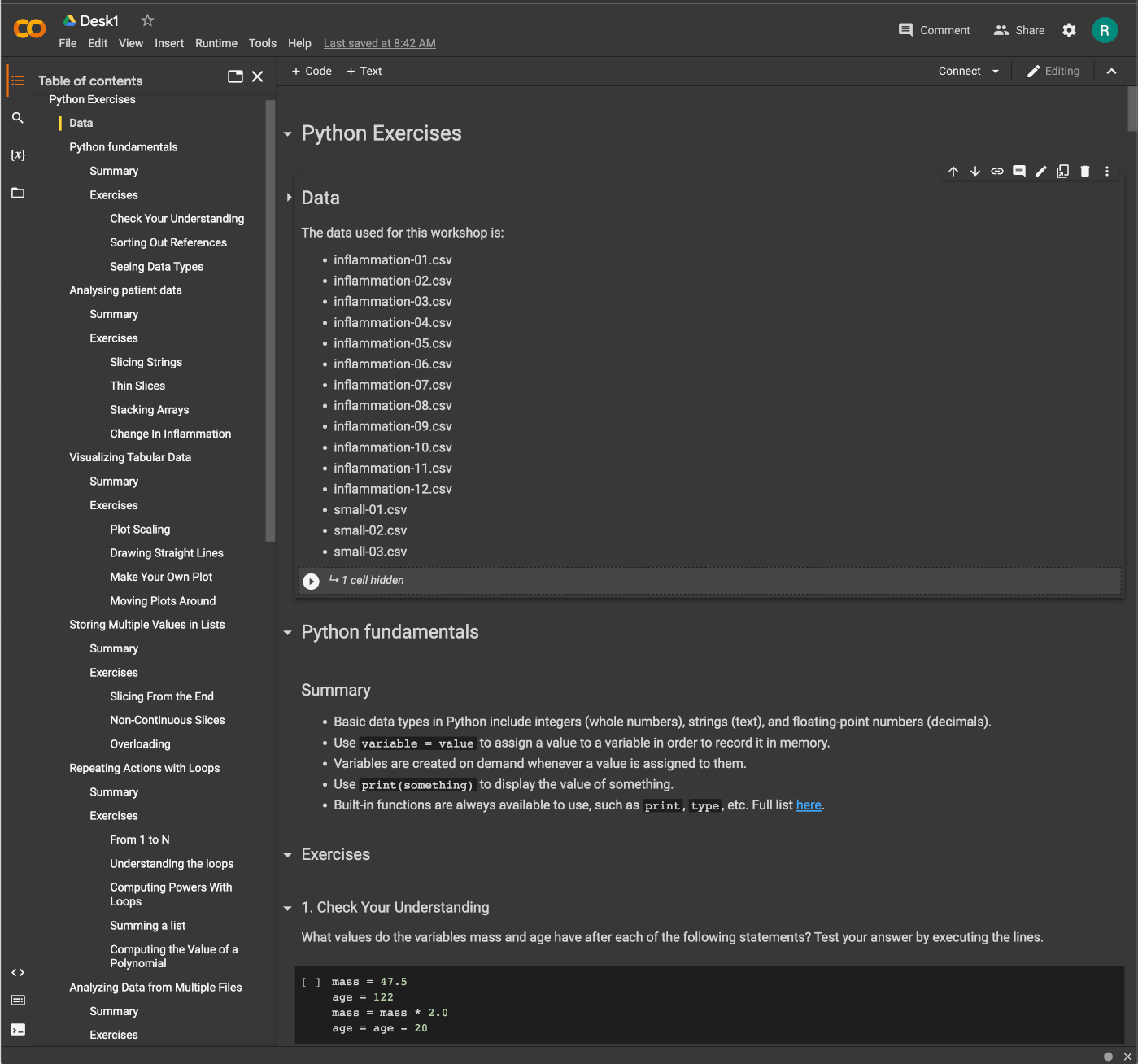}
  \caption{Google Colab notebook used for the "Python for Beginners" workshop}
  \label{fig:Colab}
\end{figure}

The workshop exercises were converted into a Google Colab notebook, one for each desk, as shown in Fig.~\ref{fig:Colab}

The Gather platform itself, from previous experiences, has occasional technical issues and so as a backup a "Zoom avatar" was also deployed to connect the Gather space to Zoom. The "Zoom avatar", ZoomBot in Fig.~\ref{fig:GatherLab} was a separate user login simultaneously hosting a regular Zoom meeting and screensharing the Gather space. This loses the ability for the participant to communicate freely with everyone in the Gather space but still allowed the participant to follow the lecture and workshop material if they found Gather was not working effectively on their system.
\pagebreak 
\section{CHALLENGES OF ONLINE TECHNICAL SKILLS TRAINING FOR THE RESEARCH COMMUNITY}
\label{section:challenges}

As highlighted in the introduction, the majority of tools adopted for online teaching and training within the research community, and indeed more widely, since the emergence of COVID-19, are best suited to a one-to-many, broadcast style of teaching. This is the approach you would get within a traditional lecture theatre/classroom setting with a lecturer/teacher talking to a group of students. Students are free to ask questions but the primary information flow is one-way. This works well for some material. However, for practical, technical content, this is generally not the most suitable teaching approach. Technical or scientific subjects use a combination of different teaching environments but practical skills such as programming are often taught within a lab environment. An instructor/lecturer will generally introduce an exercise, key skill or learning objective and learners will then work on an exercise with the ability to get help as required. Information flow is much more {\it ad hoc} and it is generally two way - a learner will explain what they've done, where they've got stuck or what they don't understand and a conversation/discussion may result to help ensure that the desired skills are understood.

The COVID-19 pandemic has resulted in an extremely rapid development of online communication tools. While such tools were widely available prior to the pandemic, there has been a phenomenal increase in their use and software providers have worked hard to address the problems that became increasingly obvious with existing tools once they were being used extensively on a daily basis by very large numbers of users. This has resulted in a wide range of easy to use, reliable video conferencing tools that have made online meetings the mainstay of many business interactions over the last two and a half years.

Naturally, training has continued during this period and training courses and workshops have been using these platforms to provide online courses. For technical training, the aspects that would previously have been carried out in a lab environment have been more challenging to support than those undertaken in a lecture theatre. Within the Carpentries training community and the university teaching environment that the authors are familiar with, there have been various attempts to use different tools and approaches to support lab-style teaching. Some of the key challenges faced in an online environment include:

\begin{itemize}
\item \textbf{Lack of learner engagement / participation}
\item \textbf{Not being able to ask a neighbour for assistance} or to discuss aspects of the work with them (when not in an assessment/examination scenario!)
\item \textbf{Being able to support multiple helpers/teaching assistants} each helping a student at the same time
\item \textbf{Gaining visual feedback from the room} - how many people have finished an exercise, how many are still actively typing away trying to complete it
\item \textbf{Resolving technical setup issues} - is everyone using the correct environment with the required software/tools?
\end{itemize}

\noindent We now explore each of these challenges in a little more detail.

\noindent  \textbf{Learner engagement:} When learning a new skill, especially in an area that you may not have prior familiarity with, it's easy to get lost and require assistance. It's also easy to feel like you're the only member of the class group who is stuck on a given problem and learners may feel uncomfortable about highlighting this in front of the rest of the group. In an in-person classroom/lab situation, a learner can raise their hand and the teacher/instructor or a helper will come over and assist. The rest of the group will continue working, or take a short pause if they are awaiting further information. Online video conferencing style platforms make this approach impractical. The focus of all participants is on the individual speaking at any given point in time. This can be intimidating or make learners feel uncomfortable if, for instance, they're raising an issue that they think may be ``obvious'' to others. Breakout rooms provide an option in such circumstances but they require the instructor or person helping to actively leave the main call to move into a breakout space, leaving them disconnected from the main space for some period of time. In such circumstances it can be easy to lose track of time if helping to address an issue is a lengthy process. The challenges of raising/handling  questions in an online environment mean that engagement of learners is often very significantly lower in online training than in-person training.

\noindent \textbf{Collaborating with or getting assistance from a neighbour:} In a regular, in-person teaching environment, learners can easily ask a neighbour if they're stuck with a problem. This may be much easier in the case of a large class and such peer support networks can form an important part of classroom or lab-based training. In a regular video conferencing call environment, there is no direct equivalent to this although direct, person-to-person, text-based chat goes some way towards providing this. Breakout rooms also offer some opportunity for peer-based technical assistance to be provided but at the expense of being disconnected from the rest of the participants for some period of time and potentially missing key information.

\noindent \textbf{Supporting multiple helpers:} In a classroom setting, a number of helpers can be available and can move round a class and each be helping someone at any point in time. The focus on a single speaker in online video conferencing platforms means that this is not possible. If a helper is talking to a learner, this will be the main communication taking place within the call and all other participants will be listening in.

\noindent \textbf{Gaining visual feedback:} This is a general challenge - when standing in a room of people, it's generally possible to gauge whether everyone is working on something, if they are stuck, or if they have finished a task. If lots of people are asking each other questions, it may be that something was unclear. If everyone is looking bored, it may be that they finished a required task some time ago and are now waiting for something else to do. Getting this kind of visual feedback from a grid of faces on a screen is much more difficult. This is made more challenging by the fact that many attendees opt to work with their cameras switched off when attending training sessions meaning that it's not possible to get any visual feedback.

\noindent \textbf{Resolving technical issues:} Technical issues can be a major problem, especially in a training environment where learners come along with their own laptop computers which all have different operating systems, software installations, etc. Even where screen sharing is an option and a learner is happy to share their screen, resolving problems online can be much slower. It's often necessary to direct the person sharing their screen to type specific commands. If they're not familiar with the commands, this can require more detailed explanation and the helper is much more reliant on gaining information from the learner. This is challenging if someone is not familiar with technical terms and infrastructure and significantly more time is required to get information about, and trial solutions to the problem. Looking more generally at the differences between Gather and standard video conferencing platforms, there are various other technical aspects that affect the user experience in different ways and these are detailed in the discussion towards the end of Section~\ref{section:discussion}.

\section {FEEDBACK}
The workshop was held on May 17-18, 2022 over two sessions running from 9am - 1pm BST following the material in Ref.~\cite{github}. Registration was capped at 60 participants, limited by the number of workstations in the virtual computer room. Despite registrations reaching capacity within a few weeks of being advertised, necessitating a waiting list, twelve people turned up on both days.

\begin{figure}[ht]
  \includegraphics[width=\linewidth]{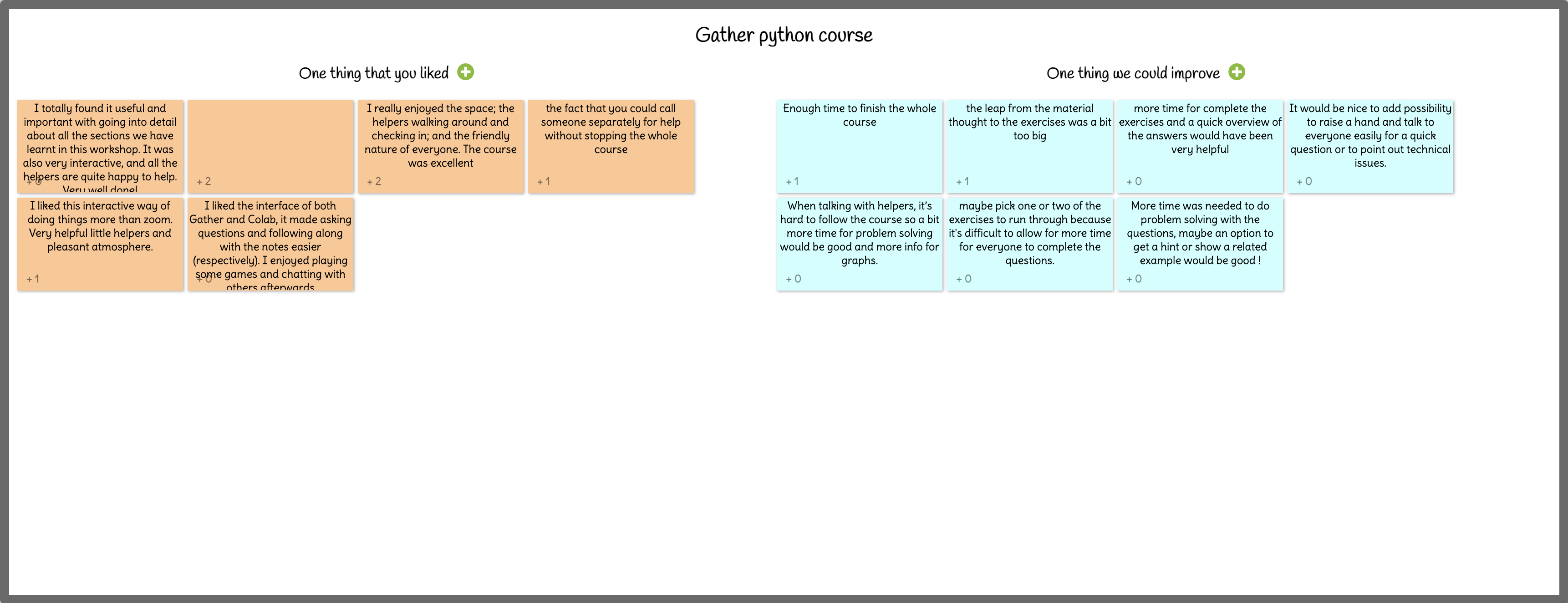}
  \caption{Feedback from the IdeaBoardz}
  \label{fig:IdeaBoardz}
\end{figure}

Evaluation of the course was carried out through the Carpentries' standard pre-workshop and post-workshop survey forms. Additional feedback on the Gather platform was solicited through use of an IdeaBoardz \cite{Ideaboardz}, an online whiteboard on which participants could leave anonymous, informal comments through virtual sticky notes. The pre-workshop survey is designed to collect information on the level of experience of the participant covering subject area, career stage and level of expertise. There were twelve respondents, a third identifying as students, and all at beginner level expressing a desire to develop their skillset. There were only two responses to the post-workshop survey which dealt mainly with the content and delivery of the course.
Further feedback, to include the use of the Gather platform, collected via an IdeaBoardz (Fig.~\ref{fig:IdeaBoardz} and Supporting Information) showed a very positive response to the space.

Impressions on the day uncovered some issues. Though the users were aware of the special features of the platform, such as bubble and private chats, they rarely used them. On the one hand, this may be related to the design of the course, because programming can be perceived as a relatively personal and individual task, and participants were focusing on their code without feeling the need to chat. On the other hand, this may also be related to the participants having come from all over the world and from different backgrounds. They were not familiar with each other, and though there was an attempted ice-breaker at the start of the session this was probably not enough to let them feel at ease with each other. On the second day of classes, however, this phenomenon eased a little. This may be because of a games session at the end of the first day, which brought the participants closer together.

From talking to the participants on the day and as reflected in the comments left on the IdeaBoardz the users felt more comfortable interacting with the other participants and the tutors. Users enjoyed the pleasant learning atmosphere - following the progress of other people, watching the tutors roaming the classroom, which is not provided by other existing online conference software. In particular, the participants appreciated being able to ask for help without delaying the learning progress of the entire course, avoiding embarrassment and making the learning experience more efficient.

\section{DISCUSSION}
\label{section:discussion}
A particular problem we have observed when running workshops online is a significant number of ``no shows''. These are people who take the time to sign up but then fail to attend and don't email to cancel. It is clear that some people will always be unable to attend a workshop at the last minute and, during COVID, we may have expected a larger number of no shows due to illness. However, we have been seeing no show rates for regular online training courses of anywhere between 40\% and 60\%. Anecdotal evidence from discussions with others undertaking similar training workshops suggests that this is a common challenge and numbers are similar in all cases. No shows at in-person workshops seem to be significantly less than this. Is this perhaps because when dealing with an online event, it feels less ``real'' and potentially participants don't see any issue with simply not attending? The commitment to register for an online, free of charge, workshop is low, whereas an in-person workshop requires plans to travel somewhere or visit a different location and is likely to require more thought prior to registration.

Nonetheless, at the workshop being described here, we observed an even higher no show rate with only around a quarter of the registered participants turning up for each day of the workshop. There are various possible causes for this but the course team wondered whether the use of Gather put some people off once they looked at it in advance of the workshop. Perhaps their computer wasn't powerful enough to run Gather reliably, or perhaps they were concerned that there might be more of a requirement to interact with the other participants than just sit and watch someone describing new topics.

Overall, the technical stack behaved just as had been hoped. There were a few audio-visual issues but these were all sorted out except in one case where the attendee may have just given up rather than troubleshoot. Such audio-visual issues have been frequently encountered in our experiences, not just with the Gather platform, and are anticipated such that we always provide an orientation or on-boarding session before the event, especially to deal with these problems. However, few people take advantage of these sessions. 

From the tutors' perspective the tech stack served its purpose well, presenting some unique advantages over Zoom-based classes, and even over traditional, in-person formats. Firstly, unlike Zoom, the spatial layout of the 2D Gather classroom faithfully reproduced in-person physical organisation, allowing tutors to quickly identify individuals who needed help (signified by them using the `raised hand' emoji), and giving students instantaneous visual feedback that a tutor was on the way without interrupting the presenter. Secondly, the mute and bubble features of Gather meant it was easy to help individuals in real-time without interrupting or distracting the presenter (as can happen in in-person classes), but also without removing the student entirely from the wider classroom context (as with Zoom breakout rooms). This minimised disruptions to the wider class, whilst also allowing struggling students to keep pace with the presenter, instead of falling behind and being left to catch up quickly in a break. Finally, the use of shareable links meant it was easy for tutors to identify and enter students' notebooks from their own laptop, read their code and any relevant errors, and provide real-time help, rather than resorting to verbal descriptions and screen-sharing. If anything this was slightly better than the in-person experience, as a tutor and student could both edit the notebook simultaneously without the awkwardness of sharing a keyboard.

The authors have both taught and attended other courses online. It is clear that the use of Gather presents a very different environment, from both a pedagogical and technical perspective, and that it also offers a number of opportunities, addressing some of the challenges described in Section~\ref{section:challenges} in novel and interesting ways. From a technical perspective, where stable, high-bandwidth Internet connectivity is available, modern video conferencing platforms offer the scope for very high quality, reliable, video and audio connections between remotely located individuals. Many widely used video conferencing platforms now provide their own client application that a user must install on their system. This helps to ensure both improved quality and reliability when compared to browser-based video conferencing by helping to avoid the differences that can exist across the many different versions of different browsers that may be in use on the systems of call participants. It also offers more flexibility in terms of user interface design and application functionality than may be possible within a browser environment. However, as the number of participants in a call begins to increase, the user experience can reduce. Not because of video or audio quality issues but simply because of the challenge of how to display video streams from a larger number of call participants in a limited set of display ``real estate''. Video windows may be made increasingly small as more participants join a call, resulting in a number of small video windows on the screen, which can be distracting and confusing. Alternatively, applications may choose to display a subset of call participants, perhaps selected based on who spoke most recently. None of these options provide an ideal technical solution when compared to the benefits of sitting in a room, in person, meeting or working with a group of people. A virtual environment such as Gather offers the promise of addressing some of these challenges and moving closer to the type of engagement and interaction that is possible through in-person training activities.

In other online training courses, using existing video conferencing platforms, that the authors have been involved with, a number of approaches have been trialled to generate more interest and engagement on the part of the attendees. This has included extensive use of chat, emoji reactions and breakout groups, often creating groups at the start of a course and having the participants use the same groups throughout with the aim of building a small community of learners and generating more engagement between group members. This process often also includes small icebreaker exercises to get group members talking and sometimes a dedicated group helper who can answer questions, develop conversations and generally help learners to get more out of their workshop experience. Such approaches do help to increase engagement. They also help to make participants feel more connected to the learning process. At the same time, the way support for these approaches is implemented within video conferencing platforms also has the effect of creating an element of disconnection between groups of training participants. When participants move to a breakout space, this generally means that they no longer have any awareness or visibility of other course participants beyond the members of their group. This affects the dynamic of the training environment but it is accepted that this is probably the only realistic technical approach in the context of existing video conferencing environments.

In Gather, the completely different nature of the environment means that there is far more scope to provide different forms of interaction. Course participants are represented as small avatars within a two dimensional visual environment displayed within their web browser. A set of several pre-prepared environments, designed to represent particular spaces (e.g. a conference hall or a classroom) are provided, and custom environments (as used in this course) can also be developed. Participants move their avatar around within the environment, potentially moving between different rooms and spaces. Other participants are also shown on the screen and when two or more people come near to each other, they each see each other's video and hear their audio. This aims to provide an experience that is much closer to that of in-person interactions. In a training context, there are a number of benefits from this. For example, the spatial chat functionality allowing students and tutors to join and leave discussions by moving their avatar around the screen is a significant improvement over breakout rooms - everyone can choose when and where to move, so students may feel less embarrassed about raising issues because they don't have to alert the whole classroom. The emoji reactions are visible as pop-ups over an avatar's head so are visible to the whole room, so tutors can easily see who is stuck and go to help them. Gather still has a chat facility that can be used if a participant does not want to turn on their microphone, but we found they were much happier to turn on both microphone and camera when in a small group discussion rather than in front of the whole class. Overall, the experience for instructors was significantly better than one-to-many video conferencing style teaching, due to the increased engagement and ease of feedback from the class. Despite the many benefits of the platform, we did observe some minor challenges. While video connections are good, the quality in the small video windows that appear overlaid on the visual environment is often not up to the level one might expect in a standard video conferencing application. For individuals unfamiliar with the platform, the technical issues relating to who can hear who, and when, also make things a little more difficult to manage. This is a function of the increased flexibility provided by Gather but it does demonstrate that this increased flexibility can provide additional complexity for users who have become familiar and comfortable with existing video conferencing platforms. The browser-based nature of Gather and other such web-based collaborative environments does provide the potential for a low barrier to accessing the service without the need to install any specific applications -- this is a significant benefit to users. These platforms take advantage of the latest web development capabilities to offer an exciting new approach to virtual collaboration and we see this as a very important development in helping to support the growth of online training.

\section{FUTURE OUTLOOK}
There was not a large enough participant sample size to form any definite conclusions. It is uncertain whether this was due to real or perceived problems with use of the Gather platform. However, in the early days of the COVID pandemic there were similar struggles to familiarise participants with the various online conferencing platforms but with no other alternatives, people persisted until it became commonplace. To make best use of Gather and future “interactive” platforms mainly requires a change in attitude. A willingness to try something new and be patient while ironing out bugs. Gather might be a more suitable platform for longer courses where the disadvantages of students not being familiar with all the features of the platform would be reduced and the huge benefit to community interaction would become much more relevant if a course took place over a few weeks rather than a couple of days.

As the world gradually emerges from COVID lockdowns and social distancing restrictions and many meetings go back to being in-person, there is a definite need for respect of the online format. The attendance issues we encountered reflect the perception that online meetings, particularly those without cost, are somehow inferior. Attendees appear less likely to give online meetings their time and attention. However, one of the practical lessons that the authors have learnt first-hand during the shift to online training is the major benefit that online training opportunities can offer in increasing accessibility of new skills to potential course participants from around the world. Where course participants don't have access to local training opportunities and where they are unable to travel, perhaps because of travel restrictions or other financial or personal circumstances, such as caring responsibilities for example, online training courses can make a huge difference to future career opportunities.

Gather has proved itself to be an engaging and effective way of delivering a traditional computer laboratory-based workshop in an online environment. Whether people will be willing in future to go out of their comfort zone and give it a chance is too early to say.

{\bf Acknowledgements: }Thank you to Gather for sponsorship of the Virtual Winter School on Computational Chemistry space.\\
The authors are also grateful to the Software Sustainability Institute for allowing us to experiment with this format in their 2022 Research Software Camp.
And special thanks to the participants who joined our workshop. Your willingness to try something new was much appreciated and we learned a lot from interacting with you.


\end{document}